# Extracting useful information from connected vehicle data: An empirical study of driving volatility measures and crash frequency at intersections

Mohsen Kamrani
Department of Civil & Environmental Engineering
The University of Tennessee, Knoxville, TN
mkamrani@vols.utk.edu

Ramin Arvin
Department of Civil & Environmental Engineering
The University of Tennessee, Knoxville, TN
rarvin@vols.utk.edu

Asad J. Khattak, Ph.D.
Department of Civil & Environmental Engineering
The University of Tennessee, Knoxville, TN
akhattak@utk.edu

## TRB PAPER # 18-00089

January 2018

**Extracting useful information from connected vehicle data:**

**An empirical study of driving volatility measures and crash frequency at intersections**

Mohsen Kamrani, Ramin Arvin, Asad J. Khattak
The University of Tennessee, Knoxville

**Abstract –** With the emergence of high-frequency connected and automated vehicle data, analysts have become able to extract useful information from them. To this end, the concept of "driving volatility" is defined and explored as deviation from the norm. Several measures of dispersion and variation can be computed in different ways using vehicles' instantaneous speed, acceleration, and jerk observed at intersections. This study explores different measures of volatility, representing newly available surrogate measures of safety, by combining data from the Michigan Safety Pilot Deployment of connected vehicles with crash and inventory data at several intersections. The intersection data was error-checked and verified for accuracy. Then, for each intersection, 37 different measures of volatility were calculated. These volatilities were then used to explain crash frequencies at intersection by estimating fixed and random parameter Poisson regression models. Given that volatility reflects the degree to which vehicles move, erratic movements are expected to increase crash risk. Results show that an increase in three measures of driving volatility are positively associated with higher intersection crash frequency, controlling for exposure variables and geometric features. More intersection crashes were associated with higher percentages of vehicle data points (speed & acceleration) lying beyond threshold-bands. These bands were created using mean plus two standard deviations. Furthermore, a higher magnitude of time-varying stochastic volatility of vehicle speeds when they pass through the intersection is associated with higher crash frequencies. These measures can be used to locate intersections with high driving volatilities, i.e., hot-spots where crashes are waiting to happen. Therefore, a deeper analysis of these intersections can be undertaken and proactive safety countermeasures considered at high volatility locations to enhance safety.

# INTRODUCTION

High-frequency connected vehicle (CV) data offers an opportunity to detect dispersions in vehicular speeds, accelerations, and jerks. Measures of dispersion attempt to quantify the spread of data. Commonly used dispersion measures include variance, range, minimum, and maximum values. In this paper, we expand the concept of "driving volatility," defined as deviation from the norm.

Volatility in driving reflects the degree to which a vehicle moves in three dimensions. If the vehicle's movements are erratic, then the risk of a crash is higher. Higher driving volatility is associated with higher safety risks, more fuel consumption, and increased emissions (*1*). The focus of this paper is to explore different measures of driving volatility, which have not yet been explored systematically in a spatial context.

CVs transmit high-frequency data between vehicles and road infrastructure. Widespread deployment of communication technologies has provided an unprecedented amount of data. Such "Big Data" combined with new tools can help researchers study, monitor, and evaluate transportation network performance in real-time (*2; 3*). This study takes advantage of the big data provided by the Safety Pilot Model Deployment (SPMD). SPMD is a field test in Ann Arbor, Michigan that offers detailed and relevant data demonstrating real-world vehicle-to-vehicle (V2V) and vehicle-to-infrastructure (V2I) communication. In this program, around 3000 vehicles, equipped with Dedicated Short Range Communications (DSRC) devices, communicate with roadside equipment (*4*). The SPMD test provides rich information packages transmitted as Basic Safety Messages (BSMs) through V2V and V2I communication. BSMs contains the vehicles position and motion information, their component status, and other information (*4; 5*). To explore the relationship between volatility and crash frequency, this study has created a new and unique database that integrates BSMs, crashes, and inventory data to extract critical information from large-scale BSM data.

This study defines measures to quantify the driving volatility in a spatial context. Then we explore correlations between the measures of driving volatilities and crash frequencies at 116 intersections in Ann Arbor, MI, where sufficient instrumented vehicles' movements were recorded. The objectives of the study are to:

1) Define and calculate several measures of volatility using vehicles' kinematic characteristics.
2) Identify measures of driving volatility (if any) that are strongly associated with crashes at intersections.

Given that driver behavior is the main contributing factor in crashes (*6; 7*), findings from this study are beneficial in two ways. First, they can help proactively identify locations with high levels of driving volatility but might not have many crashes as candidates for safety improvements. Second, reduction of driving volatility at high crash locations can reduce future crashes.

## LITERATURE REVIEW

There are various definitions for aggressive driving in the literature, but there is little agreement among them. In the current literature, researchers often use the term "aggressive" for describing behaviors that threaten the safety of drivers and occupants in the host vehicle and other vehicles. In the U.S., aggressive driving such as speeding, failure to yield the right of way, and reckless driving account for more than 50 percent of fatal crashes (*8*). Different definitions of aggressive driving have been presented in the literature. Lajunen et al. (*9*) defined driver aggression as "any form of driving behavior that is intended to injure or harm other road users physically or psychologically." These behaviors vary from less aggressive forms such as flashing lights, verbal threats, tailgating, and cutting other vehicles off, to more extreme behaviors such as physical attacks (*10*). When it comes to instantaneous driver behavior, aggressive driving can be described using different aspects of vehicle kinematics such as speed, acceleration, and vehicular jerk.

Many previous studies used common vehicle kinematics to quantify aggressive behavior or deviation from normal behavior (*11-13*). One of the more favorable variables for describing aggressiveness is maximum acceleration/deceleration of the vehicle. In the urban driving environment, Kim et al. (*14*) suggested the threshold of $1.47 \text{ m/s}^2$ and $2.28 \text{ m/s}^2$ for aggressive and extreme aggressive acceleration. De Vlieger (*15*) defined different thresholds for different driving styles in urban areas e.g., a range of $0.85 - 1.10 \text{ m/s}^2$ as an aggressive driving. Han et al. (*16*) quantified variations in driving behaviors under different driving conditions by providing different acceleration thresholds that vary with speed of the vehicle. Vehicular jerk, change in acceleration rate with respect to the time, is another element that can evaluate the aggressiveness of drivers. Vehicular jerk has been used to classify drivers' style of aggressiveness (*17*) by using the ratio of standard deviation to the mean of jerk within a time span or identifying accident-prone drivers (*18*). Feng et al. (*19*) showed that there are unique characteristics of the vehicular jerk in the gas pedal operations. Also, aggressive drivers are found to be associated with significantly higher values of vehicular jerk (*19*).

More recently, a new term "driver volatility" was introduced to describe the performance of driving behavior. The difference between "volatility" and "aggressiveness" terms is similar to the "crash" and "accident" (*20*). Different measures for driving volatilities have been used in the previous studies (*21; 22*). Kamrani et al. (*22*) defined volatility score as the coefficient of variation (ratio of standard deviation to the mean) of acceleration and deceleration. To the best of authors' knowledge, different measures of driving volatility have not been explored systematically, especially in the transportation context. Therefore, this study comprehensively explores several measures of driving volatility (applied to BSM data) and investigates their associations with intersection crash frequency.

## METHODOLOGY

Various instantaneous driving measures can be used to quantify driving volatilities such as acceleration, brake position, and steering angle. Volatility in instantaneous driving behavior should be measured by considering both longitudinal and lateral acceleration . Considering speed, acceleration, or jerk solely as the measure of volatility might ignore the importance of information embedded in the data. However, given a significant questionable error in the lateral acceleration data (*22*), only longitudinal acceleration, speed, and jerk are used in this paper. It should be noted that excluding lateral acceleration does not affect the results drastically for two reasons. First, the lateral acceleration is more critical where there is a noticeable amount of curvature in the travelers' trip, while the territory of the intersection in this study is limited to 150 ft from the center toward each approach. Second, in the area of an intersection, the traveled distance is short (called "passing" in this paper), and the geometry of the intersection does not allow drivers to have considerable lane changing space.

One hundred sixteen intersections were selected in the city of Ann Arbor, MI to extract BSM data consisting of speed, longitudinal acceleration (hereafter acceleration), time and geocodes. For each intersection, appropriate polygons are drawn based on 150 feet from the center of intersection toward all approaches. These polygons are used to filter the BSM data based on the longitude and latitude values available in the data. After the filtration, out of nearly 2,500,000,000 BSMs, 215,000,000 were found to be at the selected intersections. Data at this level are used for "level 1" calculations of driving volatilities (discussed later). The time and device ID variables of the BSMs are used to identify passings taken by each vehicle. Around 3,300,000 passings have been taken by more than 900 vehicles. Data from this step are used to do "level 2" calculations of driving volatilities (discussed later) at intersections. Crash and inventory data were also collected for individual intersections. The driving volatility and intersection related data are integrated to form the final dataset. The study uses rigorous modeling techniques that are suitable for the analysis of newly available volatility data.

### Measures of Driving Volatility

While some of the measures used for volatility are common, as shown in Table 1, other measures presented are relatively new in the transportation field. Variations in longitudinal control of a vehicle are reflected in speed, acceleration, and vehicular jerk. The values of vehicular speed and acceleration are available directly from BSM data while the jerk values are calculated from the acceleration values, since it is the rate of change of acceleration.

**TABLE 1 Summary of Measures for Driving Volatility Quantification**

| Measure of Driving Volatility | Formula | Applied to vehicular | | | | | | |
|---|---|---|---|---|---|---|---|---|
| | | Speed | Acceleration | | | Jerk | | |
| | | | + | - | both | + | - | both |
| *Standard Deviation* | $S_{dev} = \sqrt{\dfrac{1}{n-1}\sum_{i=1}^{n}(x_i - \bar{x})^2}$ | ✓ | | | ✓ | | | ✓ |
| *Coefficient of Variation* | $C_v = \dfrac{S_{dev}}{|\bar{x}|} * 100$ | ✓ | ✓ | ✓ | | ✓ | ✓ | |
| *Mean Absolute Deviation* | $D_{mean} = \dfrac{1}{n}\sum_{i=1}^{n}|x_i - \bar{x}|$ | ✓ | | | ✓ | | | ✓ |
| *Quartile Coefficient of Variation* | $Q_{cv} = \dfrac{Q_3 - Q_1}{Q_3 + Q_1} * 100$ | ✓ | ✓ | ✓ | | ✓ | ✓ | |
| *Percent of extreme values* | $\%T = \dfrac{c > Threshold}{n} * 100$ <br> $Threshold = \bar{x} \pm z * S_{dev}$ | ✓ | | | ✓ | | | ✓ |
| *Time-varying stochastic volatility* | $r_i = \ln\left(\dfrac{x_i}{x_{i-1}}\right)*100$ <br> $V_f = \sqrt{\dfrac{1}{n-1}\sum_{i=1}^{n}(r_i - \bar{r})^2}$ | ✓ | | | | | | |

*Standard Deviation*

A key measure for quantifying volatility is the standard deviation ($S_{dev}$) which is a simple and desirable statistic used for expressing variation in data:

$$S_{dev} = \sqrt{\frac{1}{n-1}\sum_{i=1}^{n}(x_i - \bar{x})^2} \qquad (1)$$

Where $x_i$ is the value of observation $i$, $\bar{x}$ is the mean, and *n* are the number of observations.

*Coefficient of Variation*

A basic measure of dispersion is the coefficient of variation which is obtained from the division of the standard deviation by the mean (*22; 23*), providing a relative measure of dispersion shown in Equation (2).

$$C_v = \frac{S_{dev}}{|\bar{x}|} * 100 \qquad (2)$$

Where $S_{dev}$ and $\bar{x}$ are the standard deviation and the mean respectively.

*Mean absolute deviation around central point*

This measure is defined as the average distance between each observation and the central tendency of the dataset (here mean) which is defined as (*24*):

$$D_{mean} = \frac{1}{n}\sum_{i=1}^{n}|x_i - \bar{x}|$$

(3)

Where $x_i$ is the observation $i$, $\bar{x}$ is the mean, and $n$ are the number of observations.

*Quartile Coefficient of Variation*

Another measure for describing dispersion of a dataset is the Quartile Coefficient of Variation, especially when the sample has non-normal distribution. The quartile coefficient of variation is defined as (*25*):

$$Q_{CV} = \frac{Q_3 - Q_1}{Q_3 + Q_1} * 100$$

(4)

Where $Q_1$ and $Q_3$ are the sample 25th and 75th percentiles respectively.

*Count of extreme values*

This measure captures driving volatility by counting the number of observations beyond a defined threshold-band. Equation (5) is showing the function (*26*) :

$$\%T = \frac{c > Threshold}{n} * 100$$

(5)

Where $c$ is the count of observations beyond the threshold and $n$ is the total number of observations. The threshold-band can be defined as (*26*):

$$Threshold = \bar{x} \pm z * S_{dev}$$

(6)

Where $\bar{x}$ is the mean, and $S_{dev}$ is the standard deviation; $z$ represents the distance between a mean and a point in units of standard deviations, i.e., $z = 1, 2, 3$, etc. Application of this measure takes into account the magnitude vehicular speed, when calculating volatility of acceleration (*22*). Figure 1 shows how the speed bin concept is applied to the real-world acceleration data obtained from the BSMs. Notably, the ability of a vehicle to accelerate declines with higher speeds. Therefore, instead of having a fixed pair of upper and lower bounds to count the number of acceleration and deceleration extreme points, speed bins of 5 mph are used in this study. The upper and lower bound for each bin are calculated using its mean and standard deviation. Similarly, vehicular jerk is classified based on corresponding speed bins.

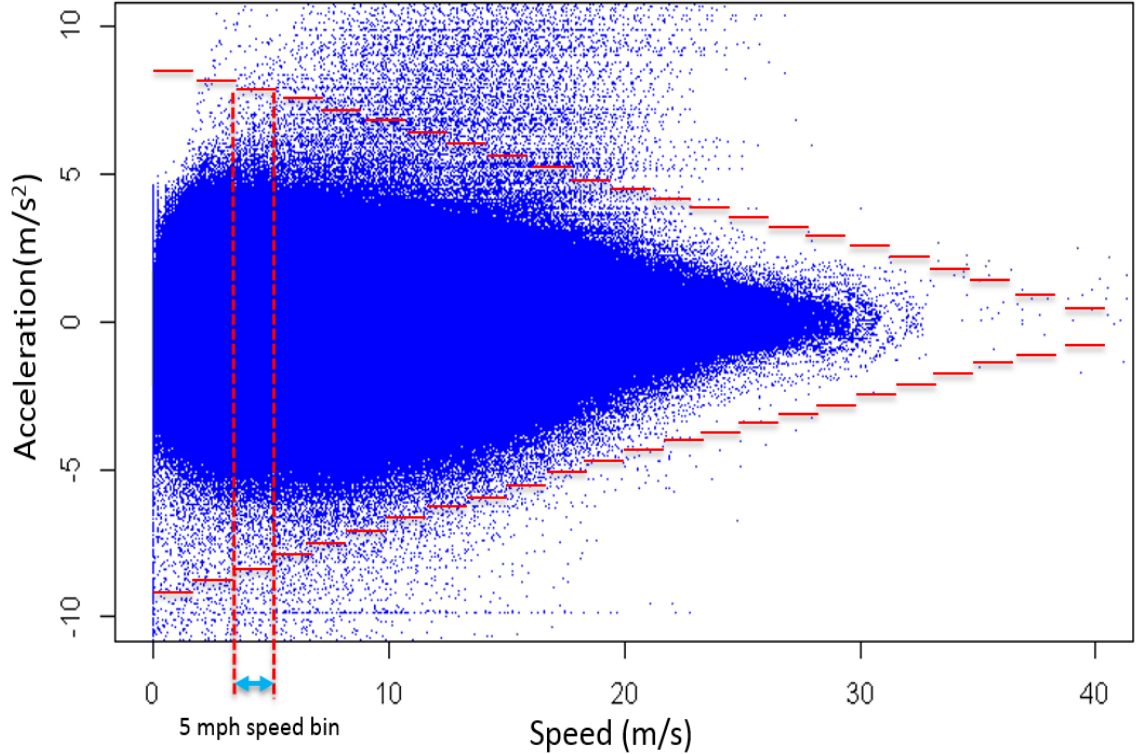

**FIGURE 1 Speed bins for calculating acceleration thresholds at various speeds using BSMs data.**

*Time-varying stochastic volatility*

The time-varying stochastic volatility which is commonly used in finance is computed by (*27; 28*):

$$V_f = \sqrt{\frac{1}{n-1}\sum_{i=1}^{n}(\mathrm{r_i} - \bar{r})^2} \quad \text{from } i = 1 \text{ to } n \tag{7}$$

Where

$$\mathrm{r_i} = \ln\left(\frac{x_i}{x_{i-1}}\right) * 100 \tag{8}$$

and $x_i$ and $x_{i-1}$ are the current and previous observations (in this study instantaneous vehicular speeds) respectively and *ln* is the natural logarithm. This measure requires positive time-series observations; therefore, it is not applicable to the acceleration and jerk values due to their negative values. Using only the positive values of acceleration and jerk will be inconsistent with the time-series nature of data required by this measure. That said, this measure is applied to speed at the vehicle passing level (level 2), which is discussed next.

**Two Levels for Calculating Volatility**

Volatility measures can be applied in two ways to obtain driving volatility at intersections as shown in Figure 2.

*Level 1 calculation of volatility*

The level 1 calculation of driving volatility disregards the individual passings (vehicles trips crossing the intersection) and treats all data for each intersection as bulk (at the aggregate level, N~215,000,000). Compared with Level 2, this calculation is simpler, easier, and faster to perform.

*Level 2 calculation of volatility*

In this method, volatility of each passing at the intersections is calculated separately. For this, the time and device ID available in BSMs are used to identify the passings. The averages of calculated volatilities for all passings are reported as measures of volatility for each intersection. Nearly 3,300,000 passings were identified for 116 intersections during the two-month period taken by around 900 unique device IDs.

**Notation of Variables**

Applying each of the measures to the speed, acceleration, and jerk at two levels results in 37 driving volatility values for each intersection. To distinguish them, a notation system is used where the volatilities have three terms in their names separated by dash "-".

- The first term is either "$L_1$" for "Level 1" or "$L_2$" for "Level 2" indicating the method of calculation.
- The second term indicates the element to which the volatility measure is applied. Since some of the measures of volatilities necessitate the separation of positive and negative values, the second term can have the following notation:
    - ***Speed:*** vehicular speed
    - ***AccDec:*** both positive and negative values of acceleration
    - ***Accel:*** positive values of acceleration
    - ***Decel:*** negative values of acceleration
    - ***Jerk:*** vehicular jerk calculated from acceleration
    - ***PosJerk:*** positive values of jerk
    - ***NegJerk:*** negative values of jerk
- The last term shows what measure was applied to obtain the volatility. For example, if **standard deviation** is applied to the **acceleration (both positive and negative values)** for individual **passings (level 2)**, the variable will be named: *"$L_2$-AccDec-$S_{dev}$"*.

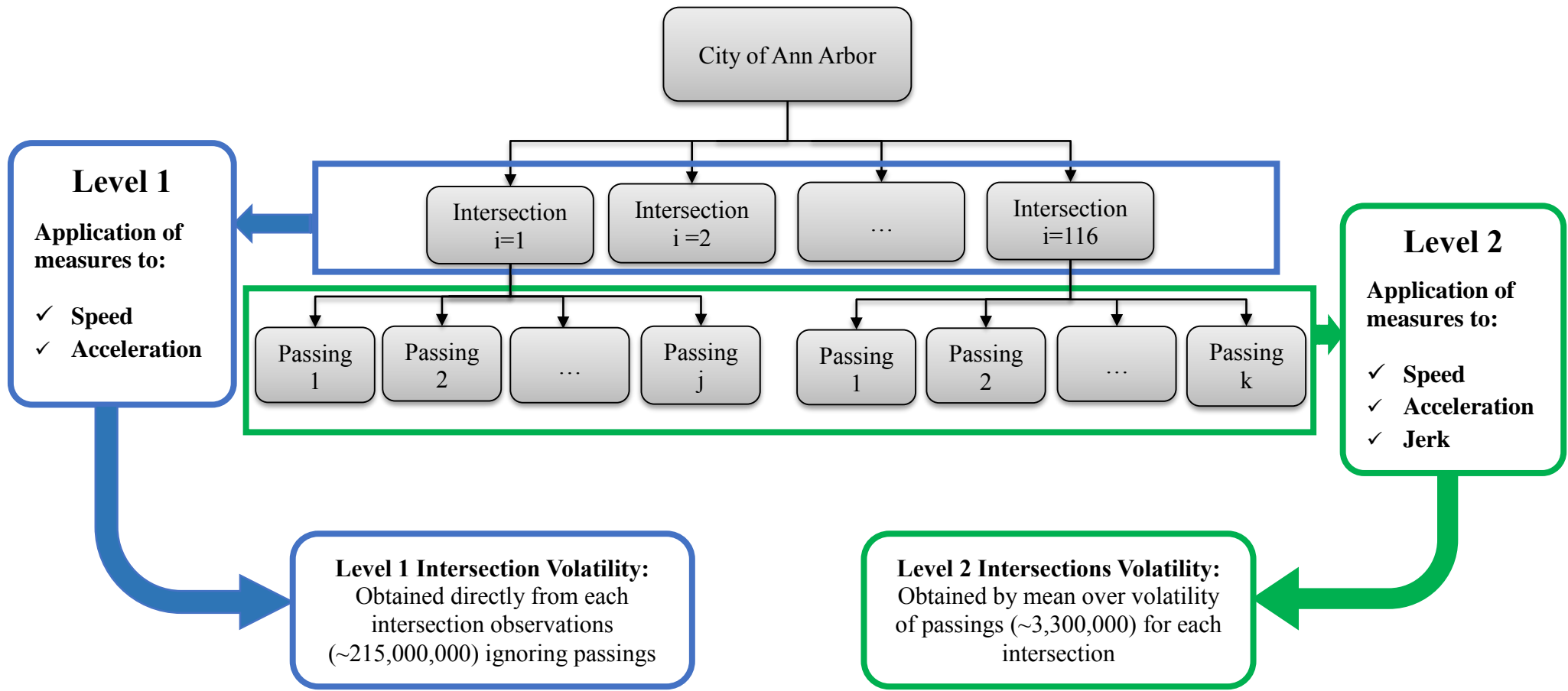

**FIGURE 2 Measures of Driving Volatility at Intersections.**

**Modeling Approach**

Count-data models are commonly used for modeling accident frequency since the number of crashes on a roadway or intersection is per unit of time and is a non-negative integer (*29*). Count data are usually modeled using Poisson or its derivatives Negative Binomial and zero-inflated models (*30; 31*). For the Poisson regression model, the probability of having *n* crashes at intersection *i* is (*32*):

$$P(n_i) = \frac{\lambda_i^{n_i} \exp(-\lambda_i)}{n_i!} \tag{9}$$

Where $P(n_i)$ is the probability of having *n* crashes at intersection i, $\lambda_i$ is the Poisson parameter for the intersection *i*. These are the expected number of crashes for the intersections in each year. In order to fit the model, $\lambda_i$ can be expressed in the logarithm form as the function of a set of independent variables (*32*):

$$\ln(\lambda_i) = \beta X_i \tag{10}$$

Where $X_i$ is a vector of explanatory variables; $\beta$ is a vector of estimated coefficients. The Poisson function defined in Equation (9) and (10) can be maximized by standard maximum likelihood procedures.

Applying Poisson regressions to the data while the mean and variance are not equal, might lead to inappropriate results. To address over-dispersion ($E(n_i) < VAR(n_i)$), or under-dispersion ($E(n_i) > VAR(n_i)$) in the data, the Negative Binomial model can be derived as:

$$\lambda_i = \exp(\beta X_i + \varepsilon_i) \tag{11}$$

Where error term, $\exp(\varepsilon_i)$, is a gamma-distributed with mean 1 and variance α. The additional term, allows variance to be different from the mean:

$$Var\ (n_i) = E(n_i) + \alpha E(n_i)^2 \tag{12}$$

Where $Var(n_i)$ and $E(n_i)$ are the variance and the expected number of crashes respectively.

Choosing between Poisson and Negative Binomial regression depends on the estimated α parameter. If α significantly does not differ from zero, Poisson regression model should be used. Otherwise, the Negative Binomial model is appropriate (*33*). Although the presence of over-dispersion can be evaluated by the mean and variance of crash data (*33*), a Lagrange multiplier (LM) can be used to statistically test the existence of overdispersion in Poisson model (*32*).

On the other hand, it is possible that associations between independent variables and the dependent variable is not consistent across all observations. Several observed and unobserved factors associated with crash frequency might lead to unobserved heterogeneity *(34-38)*. To address the heterogeneity with random parameters, using simulated maximum likelihood estimation, Greene (*32*) developed an approach to model random parameters in the Poisson model. Equation (13) indicates the formulation of estimated coefficients:

$$\beta_i = \beta + \varphi_i \tag{13}$$

Where $\varphi_i$ is a randomly distributed term with any specified distribution (e.g., normal distribution with mean zero and standard deviation σ). The Negative Binomial parameter in Equation (10) can be written as:

$$\lambda_i | \varphi_i = e^{(\beta X_i + \varepsilon_i)} \tag{14}$$

The log-likelihood function for the random-parameter model can be written as (*29*):

$$LL = \sum_i ln \int_{\varphi_i}^i g(\varphi_i) P(n_i | \varphi_i) d\varphi_i \tag{15}$$

Where g(.) is the pre-specified probability density function for $\varphi_i$. In order to maximize the log-likelihood function, a simulation-based approach using Halton draws can be used. Different studies (*39; 40*) have shown that Halton draws provide a more efficient distribution for numerical integration in comparison with random draws. Further details on random parameter models can be found in (*29*).

## DATA

The data used in this study are the result of integrating BSMs from the Michigan Safety Pilot with intersection crash and inventory data. The steps for data preparation are shown in Figure 3 (right). The BSMs data were collected, under real-world conditions, at the Ann Arbor test site by equipping around 3,000 vehicles with DSRC devices enabling them to log different variables including their instantaneous speed, acceleration heading, coordinates, etc. at usually 10 Hz. The data is accessible via ITS Public Data Hub (https://www.its.dot.gov/data/), maintained by the Federal Highway Administration under US DOT. Speed, acceleration, longitude, and latitude values of the complete two-month data (October and April 2012) were utilized in this study. The data examination and error-checking process shows high accuracy in the variables used in this study. For instance, the accuracy of the map created from BSMs shown in Figure 3 (left) is a good indication of data precision.

Intersection specific data such as the average number of crashes (2010-2014), annual average daily traffic (AADT), and speed limits for all approaches were collected. The dataset was error checked (via randomly double checking 10% of the data by a third person) and verified. The data can be obtained via Metropolitan Planning Organization website: http://semcog.org/Crash-and-Road-Data. Among intersections in the Ann Arbor area, 116 intersections are identified keeping in view that enough BSM data should be available for calculation of different measures of driving volatility. Finally, appropriate geocodes are used to filter out BSMs data for each intersection. These BSMs were used to calculate 37 different measures of driving volatilities. The final dataset was created by integrating intersection inventory data, crash data and computed driving volatilities.

## RESULTS

### Descriptive Statistics

Table 2 presents the descriptive statistics. For all intersections, the five-year mean of crashes is 7.56 with a standard deviation of 7.64. About 46% of the intersections are signalized, 40% of the intersections are 4-legged, and the rest are T-intersections. Table 2 also presents the descriptive statistics of variables calculated from BSM data i.e. measures of volatilities. Please note that the unit of analysis is the intersection.

### Correlations

Given the number of computed volatilities, correlation analysis may shed some light on relationships between crash frequency and driving volatilities (Figure 4). Bars in the figure are

**Data Integration and Processing Steps**

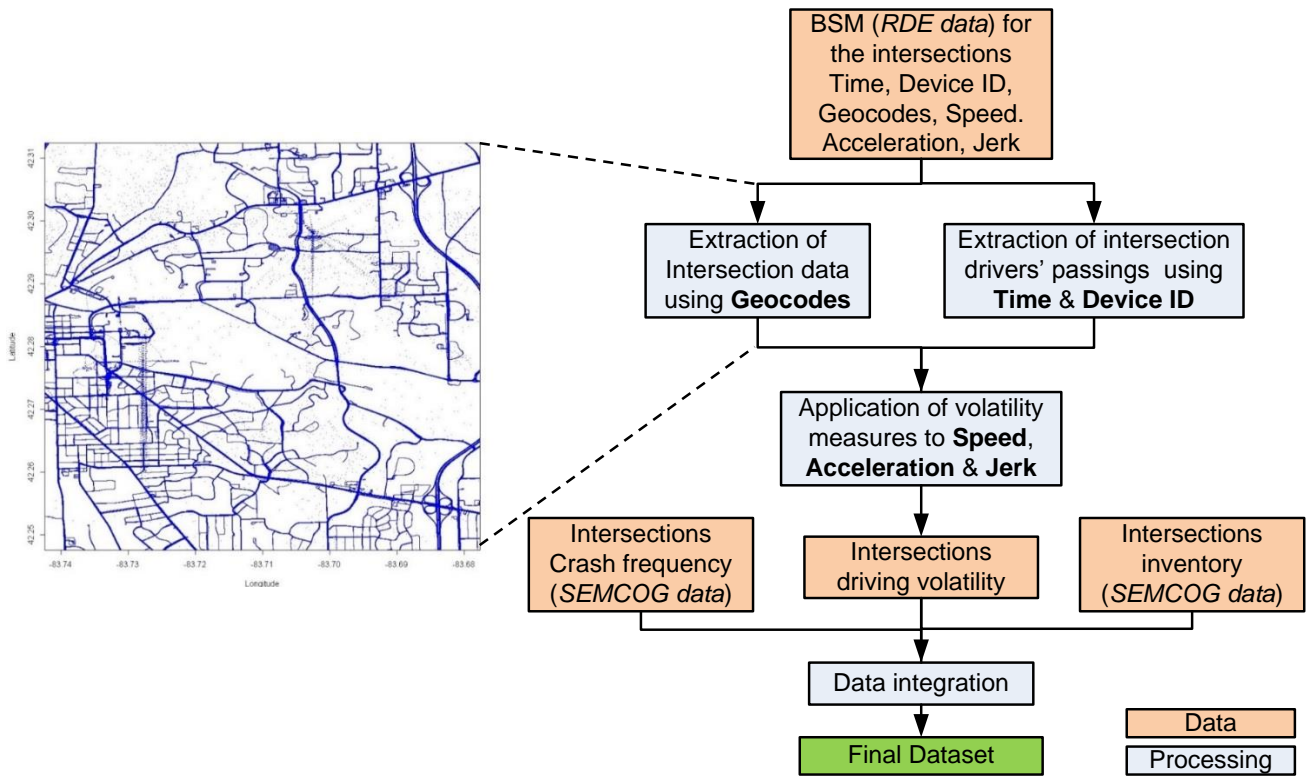

**FIGURE 3 Left: Ann Arbor map created from BSM data, Right: Data preparation**

sorted based on the value of positive correlation. Blue bars show volatilities with a positive correlation between average crashes while the red ones indicate negatively correlated volatilities. This figure was used as a guide to insert variables in the model specification and to examine their associations and improvements in model fit. As expected, there is a high level of correlation among some of the computed volatilities. For instance, two highly correlated volatilities at the bottom of the figure ($L_2$-AccDec-1$S_{dev}$ and $L_2$-AccDec-2$S_{dev}$) are calculated in a similar way, with the only difference being in the number of standard deviations from the mean. If such highly correlated variables are used simultaneously in estimation, then the model may suffer from multicollinearity. Using engineering judgment and Variance Inflation Factor (VIF>5), multicollinearity was addressed in the model specification.

**TABLE 2 Descriptive Statistics of Measures of Volatilities (n = 116)**

| Variables | Mean | $S_{dev}$ | Min | Max |
|---|---|---|---|---|
| *Intersection-related variable* | | | | |
| Average crashes (5 years) | 7.56 | 7.64 | 0 | 44 |
| AADT major road | 20805 | 8326 | 3100 | 45400 |
| AADT minor road | 9396 | 4138 | 1100 | 27400 |
| Speed limit major road | 35.34 | 7.24 | 25 | 45 |
| Speed limit minor road | 30.47 | 3.95 | 25 | 45 |
| Signalized intersection (yes = 1) | 0.46 | 0.5 | 0 | 1 |
| 4-legged intersection (yes = 1) | 0.4 | 0.49 | 0 | 1 |
| Total through lanes | 4.45 | 1.28 | 2 | 8 |
| Total left turn lanes | 1.53 | 1.32 | 0 | 6 |
| Total right turn lanes | 0.93 | 0.78 | 0 | 4 |
| *Volatility of Level 1 variables (ignoring individual vehicle passings)\** | | | | |
| $L_1$-Speed-$S_{dev}$ (m/s) | 11.35 | 2.4 | 4.92 | 16.41 |
| $L_1$-Speed-$C_v$ (%) | 45 | 16 | 13 | 71 |
| $L_1$-Speed-$Q_{cv}$ (%) | 32 | 16 | 6 | 61 |
| $L_1$-Speed-$D_{mean}$ (m/s) | 7.85 | 1.96 | 3.21 | 12.32 |
| $L_1$-Speed-%$T(1S_{dev})$ (%) | 28 | 13 | 11 | 59 |
| $L_1$-Speed-%$T(2S_{dev})$ (%) | 4 | 3 | 0 | 11 |
| $L_1$-AccDec-$S_{dev}$ (m/s²) | 0.75 | 0.17 | 0.34 | 1.43 |
| $L_1$-Accel-$C_v$ (%) | 59 | 6 | 44 | 73 |
| $L_1$-Decel-$C_v$ (%) | 65 | 9 | 51 | 103 |
| $L_1$-Accel-$Q_{cv}$ (%) | 39 | 6 | 23 | 51 |
| $L_1$-Decel-$Q_{cv}$ (%) | 44 | 7 | 23 | 59 |
| $L_1$-AccDec-$D_{mean}$ (m/s²) | 0.4 | 0.09 | 0.15 | 0.52 |
| $L_1$-AccDec-%$T(1S_{dev})$ (%) | 23 | 4 | 14 | 36 |
| $L_1$-AccDec-%$T(2S_{dev})$ (%) | 7 | 1 | 3 | 9 |
| *Volatility of Level 2 variables (averaged over passings)\** | | | | |
| $L_2$-Speed-$S_{dev}$ (m/s) | 2.02 | 0.95 | 0.41 | 5.28 |
| $L_2$-Speed-$V_f$ (%) | 2 | 2 | 0 | 6 |
| $L_2$-Speed-$C_v$ (%) | 15 | 10 | 1 | 40 |
| $L_2$-Speed-$Q_{cv}$ (%) | 10 | 7 | 1 | 26 |
| $L_2$-Speed-$D_{mean}$ (m/s) | 1.49 | 0.7 | 0.3 | 3.47 |
| $L_2$-Speed-%$T(1S_{dev})$ (%) | 34 | 2 | 29 | 39 |
| $L_2$-Speed-%$T(2S_{dev})$ (%) | 2 | 1 | 1 | 4 |
| $L_2$-AccDec-$S_{dev}$ (m/s²) | 0.4 | 0.13 | 0.17 | 1.18 |
| $L_2$-Accel-$C_v$ (%) | 27 | 6 | 15 | 43 |
| $L_2$-Decel-$C_v$ (%) | 29 | 5 | 16 | 44 |
| $L_2$-Accel-$Q_{cv}$ (%) | 18 | 4 | 10 | 28 |
| $L_2$-Decel-$Q_{cv}$ (%) | 20 | 4 | 12 | 29 |
| $L_2$-AccDec-$D_{mean}$ (m/s²) | 0.16 | 0.06 | 0.05 | 0.35 |
| $L_2$-AccDec-%$T(1S_{dev})$ (%) | 36 | 4 | 27 | 49 |
| $L_2$-AccDec-%$T(2S_{dev})$ (%) | 4 | 1 | 2 | 8 |
| $L_2$-Jerk-$S_{dev}$ (m/s³) | 1.37 | 0.15 | 1.04 | 1.78 |
| $L_2$-JerkPos-$C_v$ (%) | 59 | 3 | 52 | 65 |
| $L_2$-JerkNeg-$C_v$ (%) | 59 | 3 | 52 | 64 |
| $L_2$-JerkPos-$Q_{cv}$ (%) | 44 | 3 | 32 | 48 |
| $L_2$-JerkNeg-$Q_{cv}$ (%) | 44 | 3 | 32 | 47 |
| $L_2$-Jerk-$D_{mean}$ (m/s³) | 0.81 | 0.11 | 0.56 | 1.09 |
| $L_2$-Jerk-%$T(1S_{dev})$ (%) | 26 | 1 | 23 | 28 |
| $L_2$-Jerk-%$T(2S_{dev})$ (%) | 7 | 1 | 4 | 10 |

\* $L_1$: level 1 calculation; $L_2$: level 2 calculation; $S_{dev}$: standard deviation; %$T(1S_{dev})$: % of extreme points beyond mean ± one standard deviation; %$T(2S_{dev})$: % of extreme points beyond mean ± two standard deviation; $C_v$: coefficient of variation; $Q_{cv}$: quartile coefficient of variation; $D_{mean}$: mean absolute deviation; $V_f$: time-varying stochastic volatility; Accel: acceleration; Decel: deceleration; AccDec: both acceleration & deceleration; JerkPos: positive jerk; JerkNeg: negative jerk.

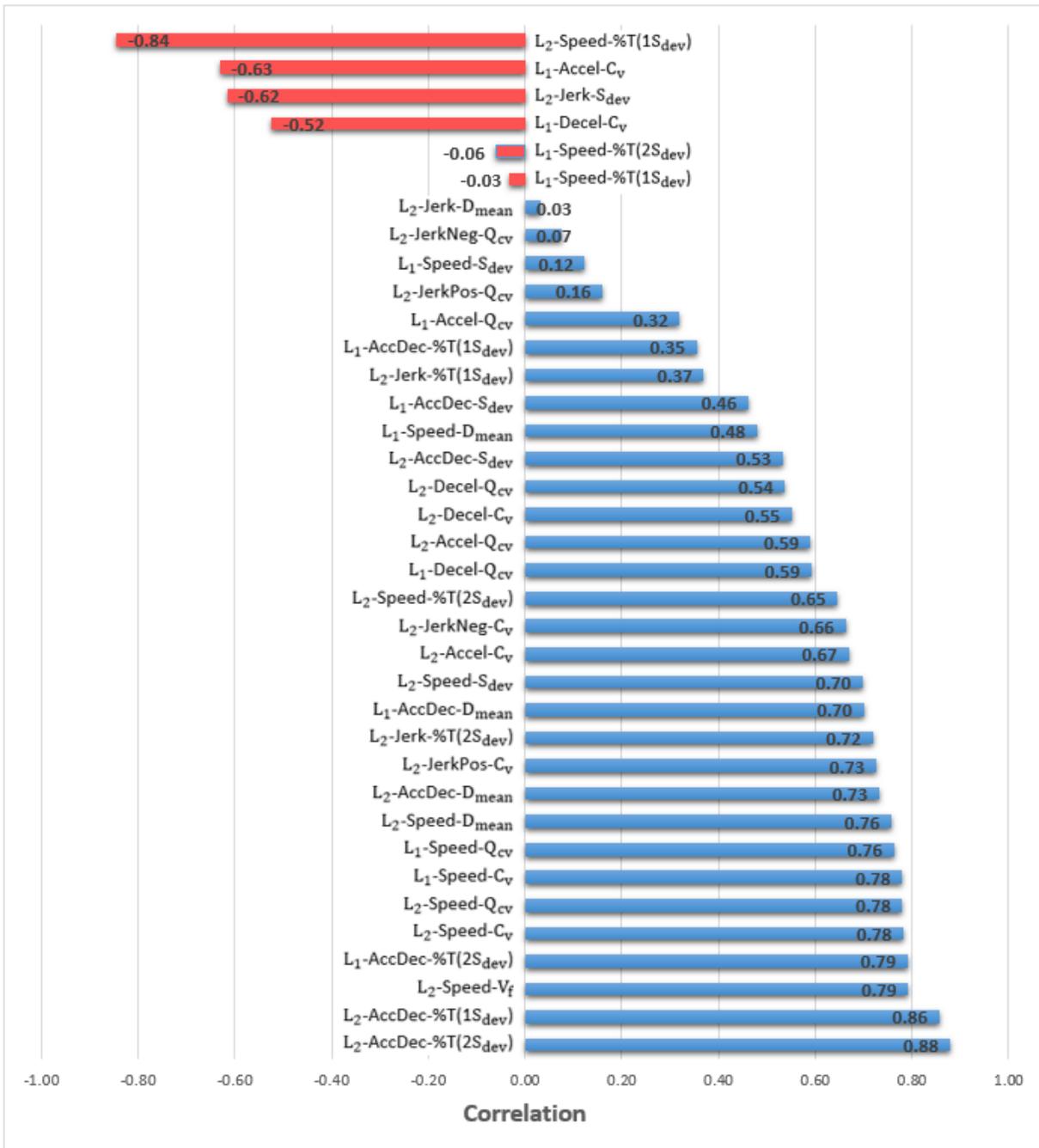

*\* $L_1$: level 1 calculation;  $L_2$: level 2 calculation; $S_{dev}$: standard deviation; %T($1S_{dev}$): % of extreme points beyond mean ± one standard deviation; %T($2S_{dev}$): % of extreme points beyond mean ± two standard deviation; $C_v$: coefficient of variation; $Q_{cv}$: quartile coefficient of variation; $D_{mean}$: mean absolute deviation; $V_f$: stochastic time-varying volatility; Accel: acceleration; Decel: deceleration; AccDec: both acceleration & deceleration; JerkPos: positive jerk; JerkNeg: negative jerk.*

**FIGURE 4 Correlations between Crash Frequency and Measures of Volatilities**

## Modeling Results and Discussion

Table 3 provides the results for fixed and random parameter Poisson regression. Fixed-parameter model is estimated for crash frequency as a function of intersection-related variables and measures of driving volatility. Starting out with intersection-related variables and keeping the significant ones in the model, measures of volatility variables were inserted in the model based on correlations from Figure 4. The models fit were compared using AIC and log-likelihood.

The random-parameter Poisson model is estimated (using simulated maximum likelihood) assuming a normal distribution for random parameters (*29*). Compared to fixed-parameter model, the random-parameter model shows a better fit based on log-likelihood, AIC, and McFadden $\rho^2$ (*33*). As Figure 5 shows, the random-parameter model outperforms the fixed-parameter in terms of crash frequency prediction.

**TABLE 3 Fixed and Random Parameters Poisson Model Results**

| Variables | Fixed Parameter | | | Random Parameter | | |
|---|---|---|---|---|---|---|
| | Estimate [a] | z value | Marginal effect | Estimate [a] | z value | Marginal effect |
| Constant | -1.497*** | -4.73 | -- | -1.852*** | -5.42 | -- |
| *Intersection-related* | | | | | | |
|    AADT major approach (1000) | 0.033*** | 7.39 | 0.25 | 0.033*** | 7.84 | 0.17 |
|    *Std. dev.* | -- | -- | -- | 0.007*** | 5.36 | -- |
|    AADT minor Approach (1000) | 0.023*** | 3.55 | 0.17 | 0.024*** | 3.70 | 0.12 |
|    Signalized intersection (yes = 1) | 0.789*** | 6.01 | 5.21 | 0.704*** | 5.77 | 3.58 |
|    Four-legged intersection (yes = 1) | 0.260** | 3.11 | 1.95 | 0.248*** | 2.93 | 1.26 |
| *Measures of volatility* [b] | | | | | | |
|    $L_1$-Speed-%$T(2S_{dev})$ | 0.050*** | 3.57 | 0.38 | 0.041*** | 2.97 | 0.21 |
|    *Std. dev.* | -- | -- | -- | 0.065*** | 8.53 | -- |
|    $L_1$-AccDec-%$T(2S_{dev})$ | 0.225*** | 4.38 | 1.70 | 0.260*** | 4.63 | 1.32 |
|    $L_2$-Speed-$V_f$ | 0.061 · | 1.92 | 0.47 | 0.109*** | 3.47 | 0.55 |
| **Summary Statistics** | | | | | | |
| AIC | 609.65 | | | 585.6 | | |
| Log-likelihood at Zero $L(0)$ | -578.32 | | | -578.32 | | |
| Log-likelihood at Convergence $L(\beta)$ | -296.83 | | | -282.81 | | |
| McFadden $\rho^2$ | 0.487 | | | 0.517 | | |
| Sample Size (N) | 116 | | | 116 | | |

[a] *Significance codes: *** 0.01%, **1%, * 5%, · 10%*
[b] *$L_1$: level 1 calculation ; $L_2$: level 2 calculation; %$T(2S_{dev})$: % of extreme points beyond mean ± two standard deviation; $V_f$: time-varying stochastic volatility; AccDec: both acceleration & deceleration.*

The marginal effects are shown in Table 3. These effects are the average increases in crash frequencies of intersections given one unit increase in the respective independent variable. For instance, an one-percent increase in the time-varying stochastic volatility of speed ($L_2$-Speed-$V_f$) is associated with a 0.55 increase in average crash frequency. That means a higher magnitude of time-varying stochastic volatility of vehicle speeds when they pass through the intersection is associated with higher crash frequencies, as expected. In addition, more intersection crashes were associated with higher percentages of vehicle data points (speed & acceleration) lying beyond threshold-bands created using mean plus two standard deviations at intersections ($L_1$-Speed-%$T(2S_{dev})$ and $L_1$-AccDec-%$T(2S_{dev})$ variables).

Other variables which are used as controls in the model show the expected signs and magnitudes. According to Table 3, 1000 more vehicles per day on the major approach are associated with a 0.17 increase in crash frequency. As expected, the association of the minor approach AADT is less than the major approach ADDT. One-thousand more vehicles on the minor road are correlated with a 0.12 increase in crash frequency. According to the model, signalized intersections on average have 3.58 more crashes than un-signalized ones. Likewise, 4-legged intersections on average have more crashes than 3-legged intersections.

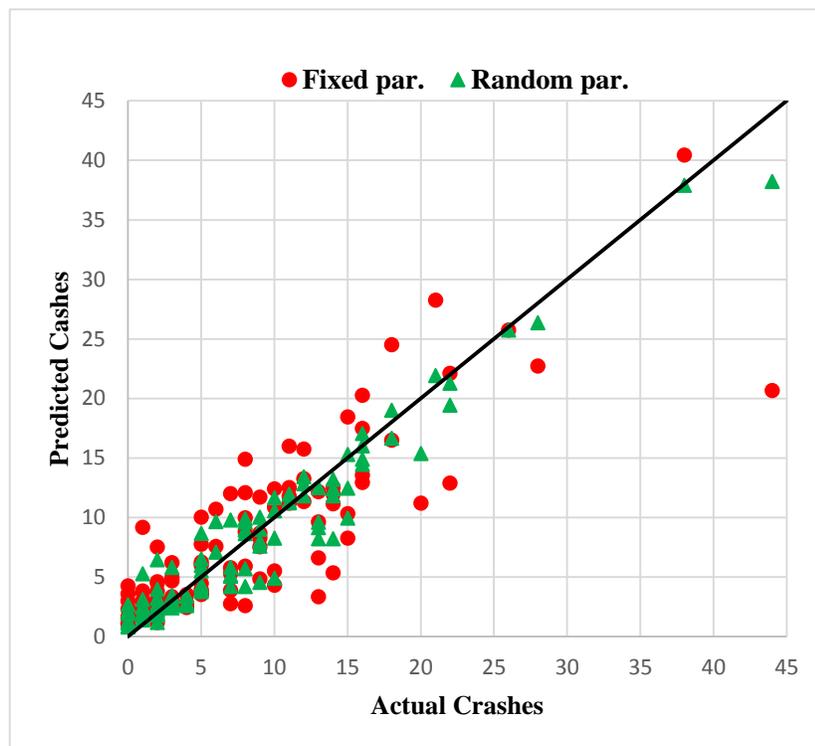

**FIGURE 5 Expected-actual number of crashes for fixed and random parameter models.**

## LIMITATIONS

The sample data used in this study does not come from representative drivers. This study did not consider volatility in the lateral direction, which could result in a sideswipe crash. Given that lane change frequency is generally relatively small at intersections, the results might not be considerably different. Furthermore, the data used in this study is the product of averaging 5-year crashes and using two-month BSMs data. In other words, a short period of instantaneous driving behavior was used to explore correlations with 5-year average crash frequencies. The authors have used all available data to make the results as accurate as possible, even though handling and processing such large-scale data was difficult. Although the data was error-checked, it is possible that some errors, made during collection of data, remain. This paper considers only crash frequency while it is worthwhile to investigate the associations of driving volatility with crash severity. Finally, it should be noted that only the means of calculated volatilities for passings (level 2 volatility) were used to model volatility at each intersection, while the between-passings variation could also be used as measures of volatility.

## CONCLUSIONS

This study discusses a way to extract useful information in the form of driving volatility from newly available BSM data. Such data are increasingly becoming available, providing a valuable resource for studying vehicle kinematics and microscopic behaviors of drivers, e.g., instantaneous vehicle speed, acceleration, and jerk. This study creates a new and unique database (BSM data integrated with crash and inventory data) and mines critical information from large-scale BSM data. More than 2,500,000,000 BSMs were processed along with crash and inventory data from 116 intersections in the city of Ann Arbor, Michigan. Volatilities of vehicles passing within 150 feet from the center of each intersection are calculated. Using nearly 215,000,000 observations for nearly 3,300,000 passings, 37 measures of driving volatility were calculated. To explore relationships between measures of driving volatility and crash frequency at intersections, rigorous statistical models were estimated. The models account for unobserved heterogeneity associated with crashes at intersections.

Three measures of driving volatilities show positive and statistically significant association with crash frequencies at the intersections. More intersection crashes are found to be associated with higher percentage of BSM data points of speed and acceleration lying beyond the threshold-bands created using mean plus two standard deviations at intersections. Furthermore, a higher magnitude of time-varying stochastic volatility of vehicle speeds when they pass through the intersection is associated with higher crash frequency. The findings are significant in the sense that they can be used to identify intersections with high levels of driving volatility. In particular, intersections where crash frequency may be low, but the volatility is high, may be good candidates for further study and future safety treatments. These are likely to be intersections where crashes are waiting to happen due to higher driving volatility. Such intersections can be proactively examined to find the causes of driving volatility to prevent crashes. Higher levels of driving volatility might be due to outdated signal timing, higher speed limits, limited line of sight, inappropriate signal timing, etc. In practice, depending on the detected reasons, proactive countermeasures can be taken to reduce drivers' volatility. In addition, appropriate alerts can be given to vehicle drivers when they are approaching locations (*41*) with a high level of driving volatility.

## ACKNOWLEDGEMENT

This paper is based upon work supported by National Science Foundation (Award number: 1538139). Additional support was provided by the US Department of Transportation through the Collaborative Sciences Center for Road Safety, a consortium led by The University of North Carolina at Chapel Hill in partnership with The University of Tennessee. Any opinions, findings, and conclusions or recommendations expressed in this paper are those of the authors and do not necessarily reflect views of the sponsors.